\documentclass[twocolumn,superscriptaddress]{revtex4}

\usepackage[latin1]{inputenc}
\usepackage[english]{babel}
\usepackage{setspace}  
\usepackage{graphicx}
\usepackage{amssymb}
\usepackage{amsmath}
\usepackage{bbold}
\usepackage{amscd}
\usepackage{natbib}
\usepackage{rotating}
\usepackage{color}
\usepackage{epstopdf}

\newcommand{\tr}{{\rm Tr\thinspace}}

\newcommand{\abs}[1]{\left\vert #1 \right\vert}
\newcommand{\expect}[1]{\left\langle{#1}\right\rangle}
\newcommand{\erf}[1]{Eq.~(\ref{#1})}
\newcommand{\var}{\mathrm{var}}

\newcommand{\T}{\mathsf{T}}

\newcommand{\F}{\mathcal{F}}

\newcommand{\BQIC}{Berkeley Center for Quantum Information and Computation, Berkeley, California 94720 USA}
\newcommand{\DeptPhys}{Department of Physics, University of California, Berkeley, California 94720 USA}
\newcommand{\DeptChem}{Department of Chemistry, University of California, Berkeley, California 94720 USA}
\newcommand{\sandia}{Sandia National Laboratories, Livermore, California 94550}

\begin{document}
\title{Qubits as spectrometers of dephasing noise}

\author{Kevin C. Young\footnote{Current affiliation: \sandia\, USA. Electronic address: \texttt{kyoung@sandia.gov}}}
\affiliation{\BQIC} 	\affiliation{\DeptPhys}

\author{K. Birgitta Whaley}
\affiliation{\BQIC} 	\affiliation{\DeptChem}

\date{\today}

%----------------------------------------------------------------
%------   Abstract
%----------------------------------------------------------------
\begin{abstract}
We present a procedure for direct characterization of the dephasing noise acting on a single qubit by making repeated measurements of the qubit coherence under suitably chosen sequences of controls.  We show that this allows a numerical reconstruction of the short time noise correlation function and that it can be combined with a series of measurements under free evolution to allow a characterization of the noise correlation function over many orders of magnitude range in timescale.  We also make an analysis of the robustness and reliability of the estimated correlation functions. Application to a simple model of two uncorrelated noise fluctuators using decoupling pulse sequences shows that the approach provides a useful route for experimental characterization of dephasing noise and its statistical properties in a variety of condensed phase and atomic systems.
\end{abstract}

\maketitle
\section{Introduction}
A key step in the design of a quantum information processing device is to gain a quantitative understanding of the decoherence-inducing 
noise processes present in the system under study.  Knowledge of the statistical properties of this noise both informs and constrains  theoretical models of the system, aiding in the design process.
In some experimental qubit realizations, the technological implementation may allow for a direct measurement of the noise.  In the superconducting flux qubit, for example, this is as simple as measuring the magnetic field fluctuations with a SQUID.  
Statistical properties of the noise may be readily computed from such a measurement and this has been very effectively used to measure the spectrum of the ubiquitous $1/f$ noise. 
Frequently, however, the noise acting on a quantum system is inaccessible to such direct measurements and the only recourse is instead to describe the 
noise indirectly through its effects on measurable quantities, such as a qubit decoherence rate.

It has been recognized for some time that appropriately selected indirect measurements may provide insight into specific features of the noise source~\cite{schoelkopf2002,faoro2004}.  For example, Schoelkopf et al.~showed in 2002 that a single qubit is a valuable resource for measuring those characteristics of the power spectrum of  an external noise source giving rise to bit flips, i.e., dissipative characteristics of the noise source \cite{schoelkopf2002}.   In many implementations, however, dephasing is the dominant decoherence mechanism \cite{tyryshkin2006}  and the techniques described in \cite{schoelkopf2002} are not directly applicable.  Subsequent work has recognized the possibility of using pulsed spectroscopies as general diagnostic tools of spectral noise, including both dissipative and dephasing properties~\cite{faoro2004,falci2004,childress2006,meriles2010}.  In particular, it has been shown that dephasing rates measured under application of dynamical decoupling pulse sequences can allow differentiation between weak and strong environmental fluctuators~\cite{nrtgli2007}.  However methods to make detailed characterization of dephasing noise that go beyond establishing such general features are still lacking.

 In this work we demonstrate that a controllable single qubit may be used as a sensitive spectrometer of dephasing noise that allows direct access to the noise correlation function.  We first present a general procedure for estimating the correlation function of such noise using sequences of coherence measurements on a single qubit undergoing free evolution, and then show that the resulting information may be extended to a signficantly larger range of time scales by use of pulse sequences that extend the coherence of a qubit by several orders of magnitude \cite{viola1999}.

A major feature of our approach is the direct nature of the procedure.  No fitting to parametric forms of correlation functions or spectral densities is required.  Finally, we emphasize that, unlike the large literature of discussions of noise sources that focus on the form of the spectral density, our discussion here is focused on extraction of the noise correlation function.  The present approach may be adapted to estimate the noise power spectral density, as we describe in Appendix~\ref{sec:freqdom}. 
However, since our reconstruction of the noise characteristics is explicitly local in the time-domain, our discussion is more naturally suited to the correlation function picture.

%----------------------------------------------------------------
%------   Model
%----------------------------------------------------------------
\section{Model}
We consider a single qubit subjected to a classical source of dephasing noise, described by the Hamiltonian
	\begin{equation} 
		H(t) = \frac{1}{2} \vec a(t)\cdot \vec\sigma + \frac{1}{2} \left(\eta_0 + \eta(t)\right) \sigma_z.
		\label{eq:hamiltonian1}
	\end{equation}
Here $\vec\sigma = (\sigma_x, \sigma_y, \sigma_z)$ are Pauli matrices and $\vec a(t)$ is a control field.
For later convenience, we have separated the second term into a constant offset field, $\eta_0$, and a zero-mean stochastic
process, $\eta(t)$. 
Such a Hamiltonian could arise, for instance, for the spin degree of freedom of an electron in a fluctuating magnetic field, a common
source of environmental noise for many atomic qubit systems, e.g., dopant spins in silicon \cite{sousa2007}. 
We assume that the qubit can be initialized in an arbitrary pure state and that it can be measured in any basis.  
We will additionally assume the noise process to be wide-sense stationary, allowing us to write the correlation
function as a function of only a single time,
\begin{equation}
 C(t) = \expect{(\eta_0 + \eta(t))(\eta_0 + \eta(0))}  = \eta_0^2 +C_\eta(t). 
 \label{eq:widestationary}
 \end{equation}
Here we have explicitly separated the full correlation function into its DC offset contribution, $\eta_0^2$, and zero-mean stochastic contribution,  $C_\eta(t) = \expect{\eta(t)\eta(0)}$.  

Experimental constraints will necessarily introduce timescales that limit what may be learned about the statistical properties of the noise from measurements on the qubit.  The proposed experiment will consist of sequences of control pulses and measurements, each of which impose a specific timescale constraint.
The longest of these timescales is the total length of time, $T$, over which the entire experiment is performed.  No matter what measurements are made during $T$, correlations long compared to this time cannot affect these measurements and are therefore not accessible.  
A second constraint is that the state of the qubit must be reinitialized between each measurement, introducing a delay $\Delta_t$ between experiments, or conversely, a ceiling on the maximum repetition rate to one measurement in time $\Delta_t$.
The shortest time scale is that of the individual measurements.  Since through the application of control pulses, the coherence of the qubit may be extended to a maximum time, $T_2$, beyond which 
the coherence has decayed to a point where it is no longer measurably different from zero, we take this coherence time as the longest time available for a measurement of the qubit.   

\begin{figure}
\begin{center}
	\includegraphics[width=\columnwidth]{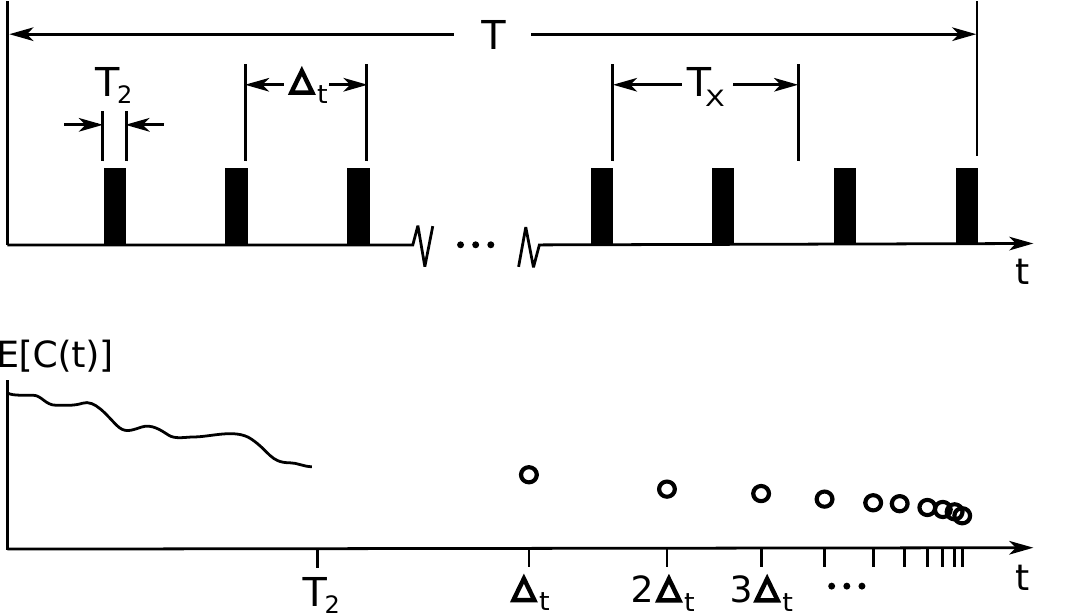}
	\caption{(a) Illustration of the various timescales involved in measurement of the noise correlation function.  
	The schematic shows a series of individual measurements (tall, black rectangles) distributed along a horizontal time axis.
	For simplicity, we have shown each measurement as lasting a time approximately equal to $T_2$, the coherence time of the qubit.
	Qubit initialization steps cause a delay $\Delta_t$ between measurements and the measurements are repeated
	until a final time, $T$.  These times define the ranges which are accessible to direct measurement of the noise correlations (see text) and $T_X$ denotes 	one of the timescales on which correlations are inaccessible to measurement.
	(b) Schematic reconstruction of a noise correlation function.  For times shorter than $T_2$, a continuous estimate is made by inversion of dynamical 		decoupling sequences (Section~\ref{sec:shorttime}).  For long times, a discrete estimate is made at times equal to integer multiples of the measurement 	time, $\Delta_t$ from measurements under free qubit evolution (Section~\ref{sec:longtime}).  The time axis is scaled logarithmically here for greater clarity.  In some systems, the short- and long-time estimates will overlap (i.e., $T_2>\Delta_t$.)}
	\label{fig:timescales}
\end{center}
\end{figure}

Taken together, these limitations provide a natural separation of the problem into short timescales within the coherence time, $t \leq T_2$, and long timescales lying in the range $\Delta_t \leq t \leq T$.  Noise correlations occurring on timescales intermediate
between $T_2$ and $\Delta_t$ are inaccessible to measurement if $T_2<\Delta_t$, as illustrated in Fig.~\ref{fig:timescales}. This puts limitations on the conversion of the directly determined noise correlations to a noise spectral density (see Appendix) and clearly implies that it is advantageous to make the measurement repetition time as short as possible.

In the following section, we construct the experimental procedures necessary to extract the correlation function on i) times long compared to the measurement separation time, $\Delta_t$, and ii) times short compared to the maximum coherence time, $T_2$.  While both methods yield direct estimates of the correlation function, the two approaches differ in the type and sequence of both control pulses and qubit measurements that are employed.    

%----------------------------------------------------------------
%------   Long-time correlations
%----------------------------------------------------------------
\section{Long-time correlations}
\label{sec:longtime}
To estimate the correlation function on timescales longer than the inter-measurement time, $\Delta_t$,
we construct a sequence of qubit measurements following different times of free evolution, $\delta_t$, according to the following protocol:  
\begin{enumerate}
	\item initialize qubit in the +1 eigenstate of $\sigma_x$, 
	\item allow qubit to evolve for a short time $\delta_t$, 
	\item measure qubit in the basis of eigenstates of $\sigma_y$.  
\end{enumerate}
We shall refer to these as free evolution (FE) measurements.
We will assume that the $j^{\rm th}$ run of this experiment begins at time, $t_j=j\Delta_t$, i.e., the runs are equally spaced in time. This is a convenient but
not essential requirement.
During step (2) above, the qubit acquires a relative phase
	\[ \phi_j = \int_{t_j}^{t_j + \delta_t} \!\!\!\left(\eta_0+\eta(s)\right)ds. \]
Upon measurement in the $\sigma_y$ basis, the measurement probabilities are
\begin{equation}
		 P_y^{\pm}(j)  = \frac{1}{2}\left(1\mp \sin \phi_j \right)\simeq \frac{1}{2}\left(1\mp \phi_j \right).
		\label{eq:measprob}
\end{equation}
Here we have taken a small angle approximation, which is valid if the total evolution time is chosen sufficiently small.  
The probabilities in Eq.~\eqref{eq:measprob} are approximately linear in the accumulated phase and may be used
to estimate the noise correlation function as follows.

Making a large number, $N$, of repetitions of the above procedure will yield $N$ measurement results.  We put these results in a vector,
 $\vec r$, with $r_j = \pm 1$ the result of the $j$\textsuperscript{th} measurement.  As shown in Appendix \ref{sec:longcorr}, this measurement result vector
 may be used to estimate the full noise correlation function at times $t_k = k \Delta_t$
	\begin{equation}
		\label{eqn:correlator}
		E[C(k \Delta_t)] = \frac{1}{\delta_t^2(N-k)} \sum_i^{N-k} r_i r_{i+k}. 
	\end{equation}
The expected error of this estimate will scale inversely with number of measurements performed for each interval, 
	\[ \var(E[C(t_k)]) \propto \frac{1}{N-k}. \]
So for a given $k$, this error can be reduced by increasing $N$, the total number of FE measurements. 

Note that if we were to measure in the $\sigma_x$ basis instead of $\sigma_y$, the measurement probabilities would then depend quadratically on the phase, $P_x^+(j) = \cos^2\!\phi_j$ and $P_x^-(j) = \sin^2\!\phi_j$.  Such probabilities are independent of the sign of
the acquired phase and are therefore not useful for determining the long correlations.  However, repeated  measurements in $\sigma_x$ are nevertheless able to give useful information about the zero-time correlation, $C(0) = \expect{\eta^2}$.  In particular, by making a series of $\sigma_x$ measurements and recording the values in a vector, $\vec r$,  we may obtain an estimate of the variance of $\eta$, i.e., $\expect{\eta^2} = (1-\expect{\sigma_x})/(2\delta_t^2)$.

%----------------------------------------------------------------
%------   Short time control
%----------------------------------------------------------------
\section{Short-time correlations}
\label{sec:shorttime}
Noise correlations on timescales between $t=0$ and $t=T_2$ cannot be investigated by the method described in the previous
section, as these timescales are generally shorter than the minimum time between measurements ($\Delta_t$).  For these short timescales we show instead that judiciously constructed sequences of control pulses allow us to make direct measurement of the overlap integral of the noise correlation function with a filter function that is defined in terms of the applied control field. 

%----------------------------------------------------------------
%------   Toggling Frame
%----------------------------------------------------------------
%\subsection{The toggling frame}
We begin by transforming the Hamiltonian \eqref{eq:hamiltonian1} into an interaction picture which removes the explicit dependence on the control fields:
	\[ H_T(t) = \frac{1}{2} \left(\eta_0 + \eta(t)\right) U_a^\dagger(t) \sigma_z U_a(t).\]
Here $U_a(t) = e^{ i \int_0^t \vec a(s)\cdot \vec \sigma ds  / 2 }$ is the unitary operator deriving from just the control field.
The simplest dynamical decoupling procedures typically limit the control fields to $\pi$-pulses polarized along $\sigma_x$ \cite{uhrig2007}.  Since $\sigma_x \sigma_z \sigma_x = - \sigma_z$, the Hamiltonian in the interaction picture remains proportional to $\sigma_z$ and becomes
	\begin{equation}
	\label{eq:toggling}
	 H_T(t) = \frac{1}{2}y(t) \left(\eta_0+ \eta(t)\right) \sigma_z,
	\end{equation}
where we have introduced the pulse function, $y(t)$, defined by
	\[ y(t) = \left\{\begin{array}{rl} 
				1 & \text{after even number of $\pi$-pulses,} \\ 
				-1 & \text{after odd number of $\pi$-pulses.}  
				\end{array}\right. \]
Though we do not consider it here, the effects of nonzero pulse widths may be included to
first-order through the modification,
	\begin{align*}
		y(t) 	
			&= \left\{\begin{array}{rl} 1 & \text{after even number of $\pi$-pulses,} \\ 
			 0 & \text{during application of $\pi$-pulses,} \\
			-1 & \text{after odd number of $\pi$-pulses.}  \end{array}\right.
	\end{align*}

The pulse function describes the fact that, from the perspective of the qubit, each $\pi$-pulse acts
to change the sign of the noise.  When averaged over all possible noise trajectories, $\eta(t)$, evolution under 
this Hamiltonian results in dephasing of the qubit, which may be quantified by the decay of the expectation value of the coherence, 
	${\sigma_+}=({\sigma_x} + i{\sigma_y})/2$.
Taking the average over all possible noise trajectories yields
	\begin{align}
		 \expect{\expect{\sigma_+(t)}} 
		 	\notag
			&= \expect{\tr \left( e^{i \int_0^t H_T(s) ds}\, \sigma_+\, e^{-i \int_0^t H_T(s) ds} \, \rho_0 \right)} \\
			\notag
			&= \expect{\exp\left(  i \int_0^t \left(\eta_0 + \eta(s)\right) y(s)\,ds \right)}\tr(\sigma_+\rho_0 ) \\
			\label{eq:cumulant}
			&= \exp\left(-\sum_l \chi^{(l)}(t)\right)\tr(\sigma_+\rho_0 ),
	\end{align}
where we have made use of a cumulant expansion in the last line \cite{kubo1962} and the terms $\chi^{(l)}(t)$ will be defined below.  Note the two different sources of averaging for the qubit coherence.
On the right hand side we have indicated averages over random variables, i.e., the stochastic average over all consistent trajectories of the noise term, with a single expectation value, $\expect{\cdot}$.  The double expectation, $\expect{\expect{\cdot}}$, of the coherence operator on the left hand side represents both this stochastic average over the noise realizations and the quantum average over the initial qubit states, denoted by the usual $Tr$ operation on the right hand side.   For our purposes here, we are interested in experiments in which the qubit is initialized into the $+1$ eigenstate of $\sigma_x$, a pulse sequence is applied, and the coherence is measured 
at a time $\tau$.   In this case, $\tr(\sigma_+ \rho_0)=1/2$. For notational convenience, we will drop the explicit dependance on $t$ from the cumulant expansion, assuming that it is always implicitly evaluated at $t=\tau$, at the end of the pulse sequence when the coherence has maximally refocused. 
The first term in the cumulant series, the  0$^{\rm{th}}$-order cumulant, is
	\[ \chi^{\left(0\right)} = -i \eta_0 \int_0^\tau y(t_1) dt_1. \]
This term, which is purely imaginary, represents the coherent precession of the qubit due to the offset field.   
The 0-order cumulant vanishes if we select any  of the many 
refocusing pulse sequences for which $\int_0^\tau y(s)ds = 0$.  The next term in the expansion vanishes, 
	\[ \chi^{\left(1\right)}=- i\!\int_0^\tau\expect{ \eta(t_1)} y(t_1) dt_1=0,\]
because the stochastic term, $\eta(t)$, has zero mean by construction.  In fact, all odd-order cumulants are purely imaginary and 
will vanish provided the 
unconditioned probability of a given noise realization, $P(\eta(t))$, is symmetric so that negative and positive contributions to the integral of $y(t_1)$ cancel.  
All even-order cumulants, $l=2$ and greater, are purely real and therefore contribute to decay of the coherence. These are therefore the terms that are responsible for dephasing of the qubit.  The dominant decoherence causing term in the expansion \eqref{eq:cumulant} is the $l=2$  cumulant 
	\begin{equation}
		\chi^{\left(2\right)}(t) = \int_0^\tau dt_1\int_0^\tau dt_2 \expect{\eta(t_1) \eta(t_2)}y(t_1) y(t_2).
	\label{eq:cumulantA}
	\end{equation}

\begin{figure}
\begin{center}
	\includegraphics[width=\columnwidth]{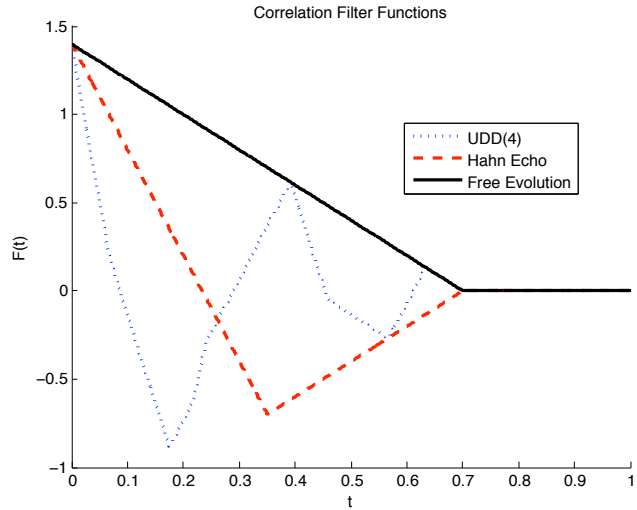}
	\caption[Example correlation filter functions.]{(Color online) Example correlation filter functions. i) (black, solid) Free evolution. ii) (red, dashed) Hahn echo. iii) (blue, dotted) 4 pulse Uhrig \cite{uhrig2007} sequence, UDD(4)}
	\label{fig:CFF}
\end{center}
\end{figure}

Making use of the assumption of wide-sense stationarity that we imposed earlier, Eq.~(\ref{eq:widestationary}), allows us to relate this expression to the stochastic part of the noise correlation function, 
$C_\eta(t_2-t_1) = \expect{\eta(t_1)\eta(t_2)}$.
By changing variables from $t_1, t_2$ to 
$u=t_2-t_1, v=t_2+t_1$, Eq.~\eqref{eq:cumulantA} may be expressed as a single-variate integral over the stochastic part of the correlation function, 
	\begin{align}
		 \chi^{(2)} 
			\label{eq:coherenceintegral}
		 	&=  \int_0^\tau du\;C_\eta(u) \mathcal{F}(u),
	\end{align}
with $\mathcal{F}(u)$ given as a second single-variate integral over a quadratic function of the control pulse sequence:
\begin{equation}
 \mathcal{F}(u) = \int_u^{2\tau-u} \!\!dv\;y\!\left(\frac{v+u}{2}\right) y\!\left(\frac{v-u}{2}\right). 
 \end{equation}
Eq.~\eqref{eq:coherenceintegral} is known as the coherence integral and defines the correlation filter function (CFF).
The CFF specifies the regions of the correlation function that contribute to dephasing under a particular pulse sequence and is only defined for 
$u\in[0,\tau]$.
Some examples of correlation filter functions resulting from different pulse sequences are shown in Fig.~\ref{fig:CFF}.

We note that discussion of spin coherence decay is usually given in terms of the noise spectral density (see e.g., \cite{ladd2009}), which is the Fourier transform of the correlation function \cite{stoica1997}.  As explained earlier, the current formulation in terms of the time-domain correlation function is preferred here since it allows for consistency with the method of elucidating long-time correlations presented in the previous section. 

%----------------------------------------------------------------
%------   Reconstruction of short time correlation function
%----------------------------------------------------------------
By taking the absolute value of Eq.~\ref{eq:cumulant}, we can remove the dependance on the 0-order cumulant (since this is purely imaginary), resulting in
	\begin{equation}
		 \abs{\expect{\expect{\sigma_+(\tau)}}} \simeq \abs{\exp\left(-\chi^{(0)}-  \chi^{(2)}\right)} = \exp\left(-\chi^{(2)}\right).
		\label{eq:experiment}
	\end{equation}	
This equation relates the CFF which is determined by and calculable from the pulse sequence, to the qubit coherence and to the noise correlation function,  Since the qubit coherence is measurable while the noise correlation function is unknown, this suggests that a direct estimate of the latter may be obtained from the former by a suitable discretization and inversion of Eq.~(\ref{eq:experiment}).  We now show how this may be done numerically.

To reconstruct the correlation function on short-timescales, one must measure the coherence decay for a number of different pulse sequences.  Each pulse sequence, $i$, determines a filter function $\F_i(t)$ as well as a coherence integral, $\chi_i^{(2)} =  \int_{0}^{\tau_i} C_\eta(t) \mathcal{F}_i(t) dt$.  We define $\abs{\expect{\expect{\sigma_+}}_i}$ as the norm of the coherence measured after the pulse sequence associated with filter function $\mathcal{F}_i(t)$.   Now because the norm of the coherence is an experimentally accessible quantity, Eq.~\ref{eq:experiment} then allows us to calculate the coherence integral $\chi_i^{(2)}$ at time $\tau$.
The procedure is then repeated for this particular pulse sequence, sufficiently many times to gather good statistics.  
By subsequently repeating the entire procedure for a large number of different pulse sequences, $\{i\}$, we will gather a set of measured coherence integrals and filter functions from which an estimate of the short-time correlation function, $E[C_\eta(t)]$, may be regressed by making use of the theory of underdetermined least-squares \cite{lawson1974}.  This results in the expression
	\begin{align}
		E[C_\eta(t)] 
			&= \sum_{ij} \chi_i^{(2)} \mathbb{F}_{ij}^+ \mathcal{F}_j(t),
			\label{eq:uls}
	\end{align}
where $\mathbb{F}_{ij} = \int_{0}^{\infty} \mathcal{F}_i(t) \mathcal{F}_j(t) dt$ is the filter function overlap matrix which may be constructed analytically for dynamical decoupling sequences or numerically for more general pulse sequences.
This set of equations is under determined because we are trying to reconstruct a continuous function by measuring a finite set of real numbers.  Consequently, there are 
in general an infinite number of possible correlation functions which are capable of reproducing the measured coherences and it is necessary to impose an optimality constraint. 
Our estimate uses the Moore-Penrose pseudoinverse, $\mathbb{F}^+$, which yields the solution with minimal Euclidean norm \cite{lawson1974}.  A derivation of Eq.~\ref{eq:uls} is given in Appendix~\ref{sec:nonorthogonal}.

%----------------------------------------------------------------
%------   Choosing pulse sequences
%----------------------------------------------------------------
\subsection*{Choosing pulse sequences}
The particular choice of pulse sequences will drastically affect the quality of the correlation function estimate, both 
by dictating the range of time over which the correlation function may be measured and by influencing the accuracy of the 
estimate within that range.  For instance, limiting oneself to a series of free evolution experiments
will only allow for an estimate of the correlation function at very short times.  Decoupling sequences greatly extend the average coherence
time, facilitating a concomitant extension of the region of the correlation function that one can estimate with this procedure.  

To quantify the sensitivity of a particular set of pulse sequences to the correlation function at a particular time, 
consider a perturbation of the noise correlation function,
	$ C_\eta(t) \rightarrow C_\eta(t) + \lambda_\tau \delta(t-\tau). $
To first order, this perturbation changes the correlation function estimate to  
	\[ E[C_\eta(t)] \rightarrow E[C_\eta(t)] + \lambda_\tau \sum_{i,j} \mathcal{F}_i(t) \mathbb{F}^+_{ij} \mathcal{F}_j(s). \]
Taking the variation of this with $\lambda$, gives a measure of the effect of the perturbation on the estimated correlation function:
	\[ \frac{\delta E[C_\eta(t)]}{\delta \lambda_\tau} = \mathcal{F}_i(\tau) \mathbb{F}^+_{ij} \mathcal{F}_j(t).\]
Squaring this quantity and integrating over $t$ provides us with a positive scalar measure of the sensitivity 
of our reconstruction to variation of the correlation function at $t=\tau$, 
	\begin{equation}
		\label{eq:quality}
		Q(\tau) = \sum_{i,j} \int_0^\infty \left(\mathcal{F}_i(\tau) \mathbb{F}^+_{ij} \mathcal{F}_j(t) \right)^2dt 
	\end{equation}
This quality function depends only on the filter functions, $\mathcal{F}_i(t)$, and the overlap matrix, $\mathbb{F}$.  
Examination of  the quality function for various sets of filter functions has empirically shown that the estimated correlation function
becomes unreliable at times for which
	\begin{equation}
		\label{eq:heuristic}
		Q(t) < \max(Q(t))/5.
	\end{equation}

%----------------------------------------------------------------
%------   Simulations
%----------------------------------------------------------------
\section{Numerical simulations}
\label{sec:simulations}
To illustrate the efficacy of our approach, we apply the procedure to a single qubit dephasing under the action of two 
mutually uncorrelated random telegraph (RT) fluctuators, using Monte Carlo techniques to simulate a 
statistically consistent noise trajectory for each measurement.
Each fluctuator is capable of existing in either of two states, $\pm\eta_i$, and will randomly transition 
from one state to the other at a rate $\gamma_i$.  To capture both the 
short- and long-time correlations that may be characterized by this spectrometry, we choose a fast, low amplitude fluctuator with parameters
$\eta_1=1, \gamma_1 = 10$
and a slow, high amplitude fluctuator with parameters $\eta_2 = 10, \gamma_2 = 0.01$.  For simplicity, we set the offset field to zero, $\eta_0=0$.
The resulting noise correlation function can be calculated from \eqref{eq:rtnprobs} to be
	$ C(t) = \eta_1^2 e^{-2 \gamma_1 \abs{t}}  + \eta_2^2 e^{-2 \gamma_2 \abs{t}}$, as shown in Appendix~\ref{sec:rtncorr}.
We simulated a series of $N=10,000$ measurements, of which the first $N_l=5000$ were free evolution decays
to measure long-time correlations, while in the last $N_s=5000$ we used dynamical decoupling to investigate the short-time
correlations.   The $j^{\rm{th}}$ measurement was assumed to begin at time  $t_j = j \Delta_t$.  
For each of the $N_l$ free evolution measurements, we simulated noise trajectories of length $t_{\rm{FE}} = 0.04$.  For each of the $N_s$ dynamical decoupling measurements a noise trajectory of length $t_{\rm{n}} = 1$ was simulated. The initial 
state of each simulated trajectory was conditioned on the final state of the previous trajectory according to Eq.~\eqref{eq:rtnprobs}.

For the short-time correlations, the result of the $i^{\rm{th}}$ measurement was simulated by first evolving the qubit under the combined action of a decoupling sequence and the $i^{\rm{th}}$ simulated noise trajectory, then by randomly selecting a measurement outcome based on probabilities calculated from the usual Born rule.  
The pulse sequences chosen to investigate the short-time coherences are given in Table~\ref{tab:sequences}.
Because of their demonstrated success \cite{biercuk2009} in extending coherence, we chose the Uhrig decoupling sequence \cite{uhrig2007} as
the basis for this simulated experiment.  
	\begin{table}[h!]
	\centering
		\begin{tabular}{|cccc|}
			\hline
			Pulse Sequence, & Time Range, & Divisions, & Repetitions\\
			\hline
			FE(1) & 0.1-0.5 & 10 & 100 \\
			UDD(2) & 0.1-0.5 & 10 & 100 \\
			UDD(3) & 0.1-0.6 & 10 & 100 \\
			UDD(4) & 0.1-0.7 & 10 & 100 \\
			UDD(5) & 0.1-0.9 & 10 & 100 \\
			\hline
		\end{tabular}
	\caption[Pulse sequences used to measure the short-time components of the noise correlation function.]
		{Pulse sequences used to measure the short-time correlations of the noise discussed in Sec.~\ref{sec:simulations}.  
		For each pulse sequence and each time step, 	one hundred noise trajectories are simulated and their effects on the qubit coherence is measured. 
		 FE = free evolution, i.e., no pulses. UDD = Uhrig dynamical decoupling sequence \cite{uhrig2007}.}
	\label{tab:sequences}
	\end{table}

\begin{figure*}[htb]
\begin{center}
	\includegraphics[width=7in]{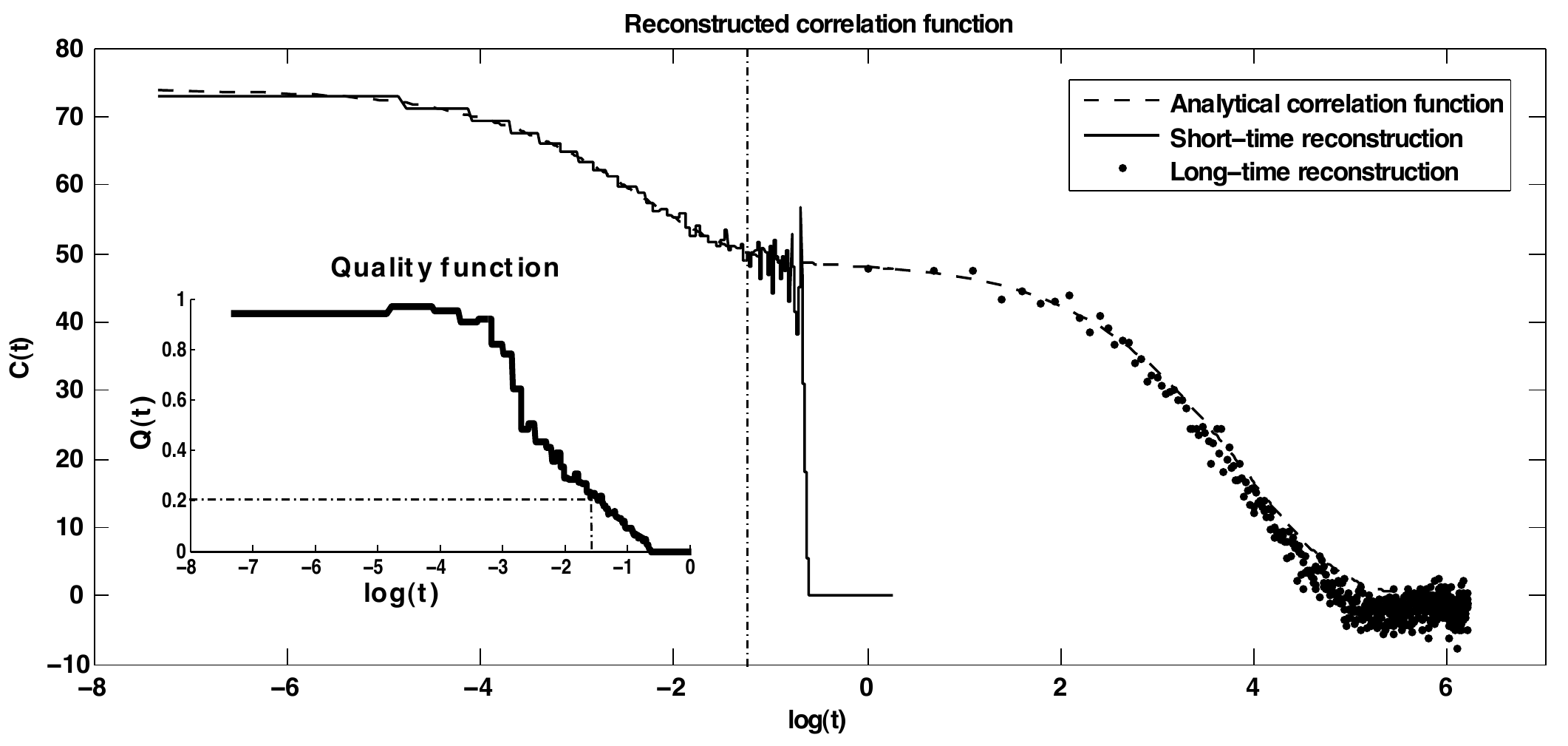}
	\caption{Combined reconstruction of the correlation function of two mutually uncorrelated RT fluctuators at both short- and long-times, obtained with use of the set of pulse sequences given in Table~\ref{tab:sequences}.
	The dashed line is the analytical correlation function, the solid line is the short-time reconstruction and the black dots represent the long-time correlations.
	 Inset is the quality function for the short-time reconstruction.  Dashed-dotted lines demarcate low quality regions.  
	 The short-time reconstruction is unreliable at times for which the quality function $Q(t) < 0.2$, corresponding to $\log(t)>-1.3$.  This unreliable portion is separated 
	 by the dot-dashed line in the main figure.}
	\label{fig:reconstruct}
\end{center}
\end{figure*}

The resulting numerically reconstructed correlation function is shown in Fig.~\ref{fig:reconstruct}, where it is compared with the exact, analytical correlation function.  The inset shows a plot of the corresponding quality function $Q(t)$. Using the heuristic Eq.~\ref{eq:heuristic}, we are able to disregard the short-time reconstruction of the noise for times $log(t)>-1.3$.   We see that the reconstructed time correlation function of the dephasing noise demonstrates remarkable overlap with the analytic correlation function at both short and long times, validating the direct reconstruction approach.

%----------------------------------------------------------------
%------   Discussion
%----------------------------------------------------------------
\section{Discussion}
\label{sec:discussion}
By inverting the conventional use of control pulse sequences, we have shown that a single qubit can be a valuable resource for direct measurement of dephasing noise.  In particular, we have demonstrated that direct reconstruction of short and long time noise correlation functions may be made using a combination of pulse sequences and free evolution measurements.  In this work we have employed dynamical decoupling pulse sequences, but the method could in principle be extended to more general sequences.  

Dephasing noise, pervasive in many quantum systems, is still relatively poorly understood.  Our direct reconstruction method is general and not dependent on any specific physical features of the probe qubit.  It may therefore be applied to any system for which dephasing is the dominant source of noise.  It is particularly well suited to the measurement of dephasing noise at interfaces, e.g., for trapped ions or for dopants in semiconductors.  One significant possible application of this approach is to the measurement of interface noise experienced by donor qubits in silicon-based devices.  As shown in \cite{tyryshkin2006}, donor qubits in silicon near an oxide interface demonstrate a marked increase in coherence time as the distance from the oxide is increased.  
Theoretical models of the noise process causing this decoherence suggest that the presence of fluctuating dangling bonds at the interface is responsible
for decoherence~\cite{sousa2007}.  However, these theoretical models require a dangling bond density which is inconsistent with the measured density~\cite{sousa2007,edwards1991}. 
The ability to make direct measurements of the
statistical properties of this noise could aid greatly in developing understanding of its microscopic origin and in construction of new theoretical models to describe the interplay of donor spins and interfaces. 

Lastly, we note that the direct reconstruction method described here is not restricted to use with a single qubit and may also be used for measurement of dephasing noise acting on ensembles of qubits.  In this situation however, unless the measurements can be spatially resolved, the reconstructed correlation function will be necessarily restricted to noise that is spatially correlated across the sample, such as that deriving from fluctuations in the applied magnetic field of a magnetic resonance experiment.  

\textit{Acknowledgements}. KY thanks Kaveh Khodjasteh and Michael Biercuk for invaluable discussion.  This work was supported by the National Security Agency under MOD713100A.  

%----------------------------------------------------------------
%------   Calculation of long-time correlation function
%----------------------------------------------------------------
\newpage
\appendix
\section{Estimate of long-time correlation function from free evolution measurements}
\label{sec:longcorr}
The correlation function at long-times is sampled by a series of free evolution measurements, yielding a results vector, $\vec r$.  We define the correlator of this result vector as 
	\begin{equation}
	C_k  =  \frac{1}{N-k} \sum_i^{N-k} r_i r_{i+k}.
		\label{eq:correlator}
	\end{equation}
The expected value of this correlator may be calculated from the measurement probabilities given in \erf{eq:measprob} according to
	\begin{align}
		\expect{C_k}
			&\simeq \frac{1}{N-k} \sum_i \expect{r_i r_{i+k}} \\
				\notag
			&\simeq \sum_{i=1}^{N-k} \sum_{m,n=\pm1} \expect{\frac{\left((-1)^m+\phi_i\right)\left((-1)^n+\phi_{i+k}\right)}{4(N-k)}}\\
				\label{eq:rescor}
			&= \frac{1}{N-k} \sum_{i=1}^{N-k}\expect{\phi_i\phi_{i+k}}.
	\end{align}
The covariance of the acquired phases may be simplified as
	\begin{align}
		\expect{\phi_i \phi_{i+k}} 
				\notag
			&=   \int_{t_i}^{t_i + \delta_t} \!\!\!\!\!\!dt_1 \int_{t_{i+k}}^{t_{i+k} + \delta_t}\!\!\!\!\!\!\!\! dt_2 \expect{(\eta_0 + \eta(t_1))\, (\eta_0 + \eta(t_2))}\\
				\notag
			&=  \delta_t^2 \eta_0^2 + \int_{t_i}^{t_i + \delta_t} \!\!\!\!\!\!dt_1 \int_{t_{i+k}}^{t_{i+k} + \delta_t}\!\!\!\!\!\!\!\! dt_2 \expect{\eta(t_1)\eta(t_2)}\\
				\label{eq:corr}
			&=  \delta_t^2\eta_0^2 + \int_{k\Delta-\delta_t}^{k\Delta+\delta_t}  C_\eta(u) f_k(u) du.
	\end{align}
In the last equality we have changed variables in the integral from $t_1$ and $t_2$ to $v = t_2+t_1$ and $u = t_2-t_1$ and 
then integrated over v.  The filter function, $f(u)$, that appears in integral is defined as
	\[ F_k(u) = \left\{ 
		\begin{array}{cl} 
		\sqrt 2 ( u - k \Delta_t + \delta_t ) & u \in [k\Delta_t - \delta_t, k\Delta_t]\\
		\sqrt 2 ( 1- u + k \Delta_t ) & u \in [k\Delta_t , k\Delta_t+\delta_t]\\
		0 & \rm{otherwise}
		 \end{array} \right. . \]  
However, from small $\delta_t$, we can assume that $C_\eta(t)$ is constant over the range $t\in[k\Delta_t-\delta_t, k\Delta_t+\delta_t]$ 
and so comes out of the integral. We can now rewrite the correlation function as
	\[ \expect{\phi_i \phi_{i+k}} = \delta_t^2 \eta_0^2 + \delta_t^2 C_\eta(k \Delta_t). \]
Combining this with Eq.~\ref{eq:rescor}, we see that
	\[ \expect{C_k}  =  \delta_t^2 \eta_0^2 + \delta_t^2 C_\eta(k \Delta_t). \]
Because the best estimate of $\expect{C_k}$ is the sample correlation, $C_k$, given in Eq.~\ref{eq:correlator}, we are left with
	\[ \delta_t^2 \eta_0^2 + \delta_t^2 C_\eta(k \Delta_t) \simeq \frac{1}{N-k} \sum_i^{N-k} r_i r_{i+k}. \]
This may be solved for the full correlation function, $C(k \Delta_t)$, as
	\[  C(k \Delta_t) \simeq \frac{1}{\delta_t^2(N-k)} \sum_i^{N-k} r_i r_{i+k}. \]
In the very-long-time limit we expect that the stochastic part becomes completely uncorrelated, 
	\[\lim_{t\rightarrow\infty} C_\eta(t) \equiv \lim_{t\rightarrow\infty} \expect{\eta(t)\eta(0)}=0.\]
The stochastic part of the correlation function may thus be recovered from the full correlation function by subtracting, i.e., 
	\[ C_\eta(t) = C(t) - \lim_{t\rightarrow\infty} C(t). \]

%----------------------------------------------------------------
%------   Calculation of RTN Correlations
%----------------------------------------------------------------
\section{Calculation of the correlation function of a random telegraph fluctuator}
\label{sec:rtncorr}
The random telegraph fluctuator is defined as a classical stochastic process taking one of two values, $\pm \eta$, 
with a flipping rate, $\gamma$.
The probability of being in the state
$\pm \eta$ at time $t$ is given by $p_\pm(t)$.  
Defining the vector $\vec p (t) = (p_+(t), p_-(t))$, the time evolution of the probabilities may be written as 
	\[ \frac{d\vec p(t)}{dt} = \left(\begin{array}{cc} -\gamma & \gamma \\ \gamma &-\gamma \end{array}\right) \vec p(t) 
		\equiv \Gamma \vec p(t). \]
This equation defines the transition rate matrix, $\Gamma$.  From this, the time evolution of the probability vector may be solved
as
	\begin{equation}
		\label{eq:rtnprobs}
		 \vec p(t)  = e^{\Gamma t} \vec p(0).
	\end{equation}
The correlation function for the noise may then be calculated to be 
	\begin{align*}
		C(t) 	&= \sum_{i,j} \eta_i P(\eta_i, t \vert \eta_j,0)  \eta_j P(\eta_j,0) \\
			&=\frac12 \vec\eta \cdot e^{\Gamma t} \cdot \vec\eta,
 	\end{align*}
which, for the transition rate matrix given above, simplifies to,
	\[ C(t) = \eta^2 e^{-2\gamma \abs{t}}.\]
Because the covariance vanishes, the correlation function of multiple, uncorrelated RT fluctuators is simply the sum of the correlation function for each fluctuator individually: 
	\[C_{ij}(t) = \expect{\eta_i(t)\eta_j(0)} = \expect{\eta_i(t)}\expect{\eta_j(0)}=0. \] 

%\subsection{Alternative calculation of RT correlation function}
%\noindent 
An alternative calculation for the random telegraph correlation function may be made as follows. In an infinitesimal time, $\delta_t$, the transition probabilities in the $i^{\rm{th}}$ fluctuator are approximately linear in the rate:
	\begin{align*}
		 P_{\text{flip}} &= \gamma \delta_t \\
		 P_{\text{no flip}} &= 1- \gamma \delta_t .
	\end{align*}
Therefore, the probability of $n$ transitions in a time interval, $\Delta_t$ is given as 
	\begin{align*}
		 P_n(\Delta_t) 
		 	&= \lim_{N\rightarrow\infty} \left(1-\frac{\gamma \Delta_t}{N}\right)^{N-n}\left(\frac{\gamma \Delta_t}{N}\right)^n \frac{N!}{n!(N-n)!} \\
			&= e^{-\gamma \Delta_t} \frac{(\gamma \Delta_t)^n}{n!} .
	\end{align*}
The last term in the first expression above is a combinatorial factor.  From this we see that the probability of an even number of transitions (which would leave the state unchanged) is 
	\[ P_e(\Delta_t) = \sum_{n \text{ even}} P_n(\Delta_t) = \frac{1}{2} \left(1+ e^{-2\gamma \Delta_t}\right), \]
while the probability of an odd number of transitions is
	\[ P_o(\Delta_t) = \sum_{n \text{ odd}} P_n(\Delta_t) = \frac{1}{2} \left(1- e^{-2\gamma \Delta_t}\right). \] 
The correlation function is then
 	\begin{align*}
		C(t) 	&= \sum_{i,j} \eta_i P(\eta_i , t | \eta_j , 0)  \eta_j P(\eta_j , 0) \\
			&= \frac{\eta^2 }{2} (2 P_e(t) - 2 P_o(t))\\
			&= \eta^2 e^{-2\gamma \abs{t}}. 
 	\end{align*}

%----------------------------------------------------------------
%------   Frequency domain filter functions
%----------------------------------------------------------------
\section{Frequency domain filter functions \label{sec:freqdom}}
The discussion in the main text for the short-time correlations may be instead expressed in terms of the spectral density.  Recall the $l=2$  cumulant is
	\begin{align*}
		 \chi^{\left(2\right)} 
		 	&= \expect{\int_0^t \eta(t_1) y(t_1) dt_1\int_0^t \eta(t_2) y(t_2) dt_2}\\
			&= \int_0^t dt_1\int_0^t dt_2 \expect{\eta(t_1-t_2) \eta(0)}y(t_1) y(t_2). \\
	\end{align*}
Using the Wiener-Khintchine theorem,  
	\[ C(t) = \int \frac{d\omega}{2\pi} e^{i\omega t} S(\omega), \]
we can rewrite the correlation function in terms of the power spectrum.  This gives
	\begin{align}
		 \chi^{\left(2\right)} 
			&= \int \frac{d\omega}{2\pi} S(\omega) \abs{\int_0^t e^{i\omega \tau} y(\tau) d\tau}^2 \notag\\
			&\equiv \int \frac{d\omega}{2\pi} S(\omega) F(\omega;t).		\label{eq:overlap}
	\end{align}
The last line above defines the filter function as the square of the Fourier transformed pulse function,
	\begin{align*}
		 F(\omega; t) 
		 	&= \abs{\int_0^t { e^{i \omega\tau} y(\tau)\, d\tau}}^2 \\
		 	&= \abs{1+ (-1)^{N+1} e^{i\omega t} + 2 \sum_{j=1}^N (-1)^j e^{i\omega \Delta_j t}}^2.
	\end{align*}
The filter functions indicate the range of frequencies of the noise power spectrum which contribute to dephasing. 
As mentioned in Sec.~\ref{sec:shorttime}, the effects of nonzero pulse widths may be included to first-order through the modification
	\begin{align*}
		y(t) 	&= (-1)^{\int_0^t a(t^\prime) dt^\prime/\pi} \\
			&= \left\{\begin{array}{rl} 1 & \text{after even number of $\pi$-pulses,} \\ 
			 0 & \text{during application of $\pi$-pulses,} \\
			-1 & \text{after odd number of $\pi$-pulses.}  \end{array}\right.
	\end{align*}
Including this modification to the pulse function changes the filter function expression to
	\[ \!\!\!\!\!\!F(\omega; t)  = \abs{1+ (-1)^{N+1} e^{i\omega t} + 2 \sum_{j=1}^N (-1)^j e^{i\omega \Delta_j t}\cos\left(\omega t_\pi /2 \right)}^2. \]

%----------------------------------------------------------------
%------   Functions in nonorthogonal bases
%----------------------------------------------------------------
\section{Estimation of correlation function by CFFs}
\label{sec:nonorthogonal}
Suppose we have chosen a large number of pulse sequences and constructed
their associated filter functions, $\mathcal{F}_i(t)$.  
As described in the main text we have experimental access to the coherence integrals,
	\[\chi^{(2)}_i = \int_0^\infty  C_\eta(t) \mathcal{F}_i(t) \,dt. \]
Because the filter functions are known in terms of the applied pulse sequences, we 
can use this integral to describe correlation function, $C_\eta(t)$.  If our set of 
filter functions were orthonormal, it would be a trivial task to expand the correlation function 
as a weighted sum of correlation filter functions, much like a Fourier series expansion.
However, this is not the case, so we instead construct a new set of orthonormal functions.  This can be done via the Gram-Schmidt 
orthogonalization procedure to yield the set
	\begin{equation}
	\label{eq:expansion}
	f_i(t) = \sum_j c_{ij} \mathcal{F}_j(t).
	\end{equation}
Properly normalized, these functions, $f_i(t)$, are orthogonal under the inner product
	\[ \left\langle f_i, f_j \right\rangle = \int_0^\infty  f_i(t) f_j( t) \,dt  = \delta_{ij}. \]
We can now expand the stochastic part of correlation function, $C_\eta(t)$, as
	\begin{align}
	 C_\eta(t)		
	 		&\simeq \sum_i  \expect{ f_i(t), C_\eta(t) } f_i( t) \notag\\
	 		&= \sum_{i,j} c_{ij}  \expect{ \mathcal{F}_j(t),  C_\eta(t) }  f_i( t) \notag\\
	 		&= \sum_{i,j} c_{ij}  \chi_j^{(2)}  \sum_k c_{ik}\mathcal{F}_k( t) \notag\\
			&= \vec\chi^\T \cdot \mathbf{c}^{\T} \!\mathbf{c} \cdot \vec{\mathcal{F}}( t). \label{eq:matrixExp}
	\end{align}
Here, $\vec\chi$ and $\vec {\mathcal{F}}$ are the vectors of measurement outcomes and filter functions, respectively, and $\mathbf{c}$ is the matrix of expansion coefficients from \eqref{eq:expansion}. We can determine the matrix $\mathbf{c}^{\T} \!\mathbf{c}$ by examining the orthogonalized filter functions
	\begin{align*}
		\expect{f_i(t),f_j(t)} &= \sum_{m,n} \expect{ c_{im} \mathcal{F}_m(t), c_{jn} \mathcal{F}_n(t)} \\
		 &= \sum_{m,n} c_{im} c_{jn} \expect{ \mathcal{F}_m(t), \mathcal{F}_n(t) } \\
		 &= \delta_{ij}.
	\end{align*}
These last two lines may be cast as a matrix equation, $\mathbb{1} = \mathbf{c} \mathbb{F} \mathbf{c}^\T$.  Here $\mathbb{F}_{ij} = \expect{\mathcal{F}_i,\mathcal{F}_j}$ is the filter overlap matrix and $\mathbb{1}$ is the identity matrix. We see then that $ \mathbf{c}^\T \mathbf{c} = \mathbb{F}^{-1}$.  However, $\mathbb{F}$ is likely to be numerically ill-conditioned, so we replace $\mathbb{F}^{-1}$ with $\mathbb{ F}^+$, the Moore-Penrose pseudoinverse of $\mathbb{F}$.  We can now rewrite \eqref{eq:matrixExp} as
	\begin{equation}
		\label{eq:estimator}
		C_\eta(t) \simeq \vec{\chi}^\T\cdot \mathbb{F}^{+}\cdot\vec{\mathcal{F}}(t).
	\end{equation}	
In reality one will only be able to perform a finite number of experiments, so this expansion is only approximate (much as a finite Fourier-expansion is only an approximation of the expanded function).  We point out that we are able to express our estimate power spectrum entirely in terms of the filter functions, their overlap integrals, and experimentally accessible parameters. It is not necessary to explicitly construct the orthogonalized filter functions defined in \eqref{eq:expansion}, and we avoid the numerical errors commonly associated \cite{bjork1994} with the Gram-Schmidt procedure.

\bibliography{thesis}
\bibliographystyle{apsrev}

\end{document}